\documentclass[twocolumn,amsmath,amssymb,nofootinbib,superscriptaddress]{revtex4}
\usepackage{graphicx}
\usepackage{verbatim}
\usepackage{amsmath,amssymb}
\begin{document}
\title{
Universal quantum magnetometry with spin states at equilibrium
}
\author{Filippo Troiani}
\email{filippo.troiani@nano.cnr.it}
\affiliation{Centro S3, CNR-Istituto di Nanoscienze, 
I-41125 Modena, Italy}
\author{Matteo G. A. Paris}
\email{matteo.paris@fisica.unimi.it}
\affiliation{Quantum Technology Lab,
Dipartimento di Fisica dell'Universit\`{a} degli Studi di
Milano, I-20133 Milano, Italy}
\affiliation{INFN, Sezione di Milano, I-20133 Milano, Italy}  
\affiliation{Department of Mathematics, Graduate School of Science, Osaka University, Toyonaka, Osaka 560-0043, Japan}                                       
\begin{abstract} 
We address metrological protocols for the estimation of the intensity and the orientation of a magnetic field, and show that quantum-enhanced precision may be achieved by probing the field with an arbitrary spin at thermal equilibrium. A general expression is derived for the ultimate achievable precision, as given by the quantum Fisher information. The optimal observable is shown to correspond to the spin projection along a temperature-dependent direction, and allows a maximally precise parameter estimation also through ensemble measurements. Finally, we prove the robustness of our scheme against deviations of the measured spin projection from optimality.
\end{abstract}
\maketitle
Quantum sensing and metrology are the arts of using quantum system to design precise estimation protocols. In order for this approach to be efficient, one needs, on the one hand, to exploit the inherent sensitivity of quantum systems to external disturbances, and, on the other, to minimize the detrimental effects of such sensitivity, e.g. decoherence and measurement back action \cite{Helstrom_1976a,Braunstein_1994a,Giovannetti_2006a,Degen_2017a}.
\par
Quantum parameter estimation is the paradigmatic problem of quantum metrology. Here, the parameter of interest is not directly measurable, 
but rather encoded into the state of a physical system, from which it 
has to be extracted by suitably designed measurement schemes. Within the framework of parameter estimation theory, one may identify the optimal observable and evaluate the corresponding sensitivity. Such ultimate quantum limit to the achievable precision is given by the quantum 
Fisher information \cite{Paris_2009a}. 
\begin{figure}
\begin{center}
\includegraphics[width=0.450\textwidth]{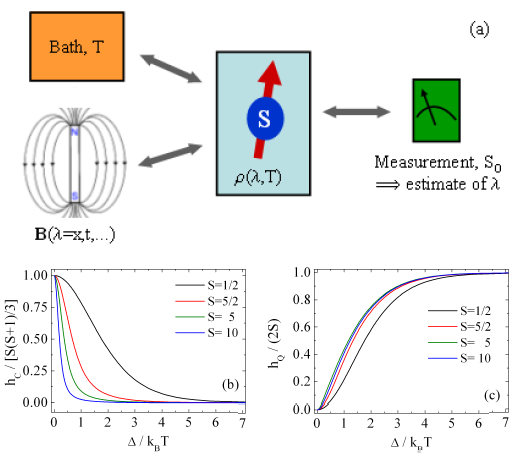}
\caption{\label{fig1}
(a) Schematics of the problem: a spin of length $S$ is coupled 
to a parameter-dependent magnetic field ${\bf B}(\lambda)$, and is
in equilibrium with a heat bath at a temperature $T$. The value 
of $\lambda$ is inferred from the measurement outcome of a spin 
observable $S_O$. Panels (b) and (c) show the temperature dependence of the 
classical ($h_C$) and quantum ($h_Q$) components of the 
quantum Fisher information, normalized respectively to $S(S+1)$ 
and $2S$. The quantum Fisher information is plotted for different 
values of the spin length (see labels).}
\end{center}
\end{figure}
Indeed, these tools have been employed to address several metrological problems, ranging from the characterization of complex environments \cite{Benedetti_2014a,Zwick_2016a} to that of quantum channels \cite{Monras_2007a,Fujiwara_2001a,Pinel_2013a} and correlations \cite{Brida_2010a,Blandino_2012a}, from the estimation of an optical phase \cite{Monras_2006a,Kacprowicz_2010a,Genoni_2011a,Spagnolo_2012a} to quantum thermometry \cite{Brunelli_2011a,Correa_2015a}. However, the explicit determination of the optimal observable, and its actual implementation 
in the lab, often remain challenging. 
\par 
A magnetic field is a typical quantity whose estimate can be obtained through a quantum probe. In fact, a variety of atomic \cite{Kitching_2011a,Kominis_2003a,Vengalattore_2003a,Maiwald_2009a} and solid-state spin systems \cite{Packard_1954a,Taylor_2008a,Rondin_2014a,Schirhagl_2014a} are currently being used in order to implement field sensors of increasing sensitivity and spatial resolution. The most widely pursued approach ultimately relies on the measurement of the quantum phase accumulated by the spin, due to its interaction with the field \cite{Degen_2017a}. These protocols typically require the coherent control and dynamics of the spin state and, in the most advanced cases, the controlled generation of entanglement in order to outperform classical devices. Systems at thermal equilibrium represent a possible alternative,  whereby the state preparation, though less general, is greatly simplified, and decoherence no longer represents a limiting factor. 
\par
In this Letter, we address the estimation of a magnetic field, obtained by performing arbitrary measurements on the equilibrium state of an arbitrary spin. We find a general expression for the quantum Fisher information, which is also given in terms of common thermodynamic quantities, such as magnetization and magnetic susceptibility. Besides, the optimal observable is shown to coincide with the spin projection along a direction that depends on temperature through a universal expression. Interestingly, the upper bound for the estimation precision is also saturated by an ensemble measurement of the optimal spin projection, which is quantified by a different figure of merit. Finally, the dependence of the Fisher information on the measured spin projection demonstrates the robustness of the estimation protocol with respect to deviations from optimality.
\par
{\it Defining the problem.---} A finite system with equispaced 
energy levels can be represented in terms of a spin of length $S$, 
placed in an external magnetic field. We 
consider the case where the field depends on the unknown 
parameter $\lambda$ (which can in principle coincide with 
spin position, time, etc.), both in intensity and orientation. 
The system Hamiltonian thus reads:
\begin{equation}
\mathcal{H} 
= 
\Delta\, (S_x \sin\theta + S_z \cos\theta)
=
\Delta\, \hat{\bf n}_Z \cdot {\bf S} 
\equiv
\Delta S_Z ,
\end{equation}
where the direction 
$ \hat{\bf n}_Z = (\sin\theta \cos\varphi,\sin\theta\sin\varphi , \cos\theta) $ 
and the energy gap $\Delta$ are known functions of $\lambda$, whose value is unknown and has to be estimated by performing (repeated) measurements of a spin observable $S_O$ [Fig. \ref{fig1}(a)]. In order to simplify the equations, and without loss of generality, we define the reference frame such that both $ \hat{\bf n}_Z $ and $ \hat{\bf n}_X \equiv \partial_\theta\hat{\bf n}_Z $ lie in the $xz$ plane (and thus $\varphi = 0$) for a generic value $\lambda$ of the unknown parameter.

The spin is in equilibrium with a heat bath at a temperature $T$. Its density operator is thus given by the following expression:
\begin{equation}
\rho_\lambda
= \!\!
\sum_{M_Z=-S}^S \frac{e^{-\delta M_Z}}{\mathcal{Z}} |M_Z \rangle\langle M_Z|
\equiv \!\!
\sum_{M_Z=-S}^S p_{M_Z} |M_Z \rangle\langle M_Z| ,
\end{equation}
where $\mathcal{Z} = \sum_{M_Z=-S}^S e^{-\Delta M_Z} $ is the partition function, $ \delta \equiv \Delta / k_B T $ is the ratio between the Hamiltonian and the thermal energy scales, and $ |M_Z \rangle = e^{-iS_y\theta} |M_z\rangle $ are the eigenstates of $ S_Z $. The density operator $\rho_\lambda$ depends on the parameter $\lambda$ through the angle $\theta$, which determines the eigenstates $|M_Z\rangle$, and through the normalized energy gap $\delta$, which determines the corresponding probabilities $p_{M_Z}$. 
\par
{\it Identifying the optimal observable. ---}
The observables that allow in principle the most precise estimate of the parameter $\lambda$ can be derived from the symmetric logarithmic derivative (SLD) $L_\lambda$, which is obtained by solving the differential equation
$\partial_\lambda \rho_\lambda = \frac{1}{2} (L_\lambda \rho_\lambda + \rho_\lambda L_\lambda)$ \cite{Helstrom_1976a,Paris_2009a}.
In order to determine the SLD corresponding to the present density operator, we compute first the derivative of $\rho_\lambda$, and then its matrix elements in the basis of the Hamiltonian eigenstates \cite{SI}. The resulting expression of the SLD reads:
\begin{eqnarray}\label{sld2}
L_\lambda \!\!
& = & 
\dot\delta \!\left( \langle S_Z \rangle - S_Z \right)
\!+\!
2\dot\theta \tanh(\delta/2) S_X\!,
\end{eqnarray}
where $ S_X \equiv \hat{\bf n}_X \cdot {\bf S} $, $\dot\theta \equiv \partial_\lambda\theta$, and $\dot\delta \equiv \partial_\lambda\delta$.

From the above equation it follows that $L_\lambda$, and thus the optimal observable, simply coincides (up to irrelevant transformations) with a spin projection $S_{O,opt}$ along a temperature-dependent direction. This is specified by the versor
$ \hat{\bf n}_{O,opt} \equiv [\sin(\theta+\phi_{opt}) , 0 , \cos(\theta+\phi_{opt})] $, lying in the $xz$ plane, with the angle $\phi_{opt}$ defined by the equation
\begin{equation}\label{eqOA}
\dot\delta \tan\phi_{opt} = - 2 \dot\theta \tanh(\delta/2).
\end{equation}
Therefore, if $k_BT$ is much larger than $\Delta$, $S_{O,opt}$ coincides with the spin projection along the field direction ($S_Z$). The same applies if the dependence of the normalized energy gap on $\lambda$ is much stronger than that of the azimuthal angle. In the opposite limits ($|\dot\delta| \ll |\dot\theta|$), $S_{O,opt}$ tends to coincide with the transverse spin component $S_X$. If the system is exactly in the ground state, a spin projection along any direction in the $xz$ plane corresponds in fact to an optimal observable (see below). 

{\it Parameter estimation with an arbitrary observable.---}
Given the expression of the SLD, one can readily derive the highest achievable precision in the estimation of the parameter $\lambda$. This is given, through the quantum Cramer-Rao bound, by the quantum Fisher information $ H = {\rm Tr} (\rho_\lambda L_\lambda^2)$ \cite{Helstrom_1976a,Paris_2009a}. In the present case, $H$ is given by \cite{SI}:
\begin{eqnarray}\label{eqQFI}
H
\!=\!
\dot\delta^2 (\langle S_Z^2 \rangle \!-\! \langle S_Z \rangle^2)
\!+\!
4 \dot\theta^2 \langle S_X^2 \rangle \tanh^2(\delta/2).
\end{eqnarray}
The expectation values that appear in the above expression are given by
$\langle S_Z \rangle 
=
(1/2)f_{1/2}-(S+1/2)f_{S+1/2}$,
$\langle S_Z^2 \rangle
=
S(S+1)+f_{1/2} \langle S_Z \rangle $
[where we have introduced the abbreviation $f_a \equiv \coth (a\delta)$], 
and 
$ \langle S_X^2 \rangle = \frac{1}{2}[S(S\!+\!1)\!-\!\langle S_Z^2 \rangle] $. 
The quantum Fisher information can thus 
be expressed in terms of thermodynamic quantities that are 
routinely measured in magnetic systems, such as the 
magnetization and the magnetic susceptibility.
\par 
Equation (\ref{eqQFI}) also shows that one may decompose the quantum Fisher 
information  
into the sum of two contributions, $H = h_C \dot\delta^2 + h_Q \dot\theta^2$. 
The term $h_C$, to which we 
refer in the following as ${\it classical}$, is determined by 
the dependence of $\delta$ on $\lambda$. This contribution is 
proportional to the fluctuations in the {\em longitudinal} spin 
projection, which result from the incoherent mixture of
the Hamiltonian eigenstates $|M_Z\rangle$. 
The term $h_Q$, hereafter referred 
to as {\it quantum}, depends on the changes of the field direction 
$\hat{\bf n}_Z$ and is proportional to the fluctuations of the 
{\em transverse} spin components. These are of genuine quantum 
origin, and are in fact present also if the system state coincides 
with an Hamiltonian eigenstate (see below). 

In order to better understand the behavior of the quantum 
Fisher information, we separately plot the classical and 
quantum contributions as functions of $\delta$, for 
a few specific values of the spin length [Fig. \ref{fig1}(b,c)]. 
The term $h_C$, which can be related to the sensing 
of the field intensity, vanishes in the limit of 
infinite $\delta$, where $\rho_\lambda$ coincides with 
the ground state $| M_Z = - S \rangle$. It increases 
monotonically for decreasing $\delta$, and achieves 
its maximum $S(S+1)/3$ for $\delta = 0$. Therefore, 
quite remarkably, one obtains a quadratic scaling of 
the quantum Fisher information with the system size 
for a highly mixed state and a seemingly classical 
term, related to thermal fluctuations of the spin 
projection $S_Z$. The quantum component 
$h_Q$, which can be related to the sensing of the 
field direction, vanishes for $\delta=0$, and approaches a 
linear dependence on $\tanh (\delta/2)$ for large values of 
the spin length. In the low-temperature limit, $\rho_\lambda$ tends to the state $M_Z=-S$ and $h_Q$ 
approaches the value of $2S$. A further discussion on the temperature dependence of the classical and quantum terms is given in Ref. \cite{SI}.
\par
The result concerning the low-temperature limit can be generalized by considering 
an arbitrary Hamiltonian eigenstate,
$\rho_\lambda = |M_Z\rangle\langle M_Z|$. In such a case, the quantum Fisher 
information reads:
\begin{eqnarray}\label{qfizt}
H
= 2 \dot\theta^2 \, [S(S+1)-M_Z^2] .
\end{eqnarray}
The highest achievable precision in the parameter 
estimation thus scales linearly with the system 
size ($S$) in the case of a coherent state ($M_Z=\pm S$), 
and quadratically for $M_Z=0$. Therefore, quite counterintuitively, a state 
characterized by a vanishing expectation value of all 
spin projections ($\langle {\bf S} \rangle = 0$) presents 
the strongest dependence on the orientation of the 
quantization axis. On the other hand, one should consider 
that the spin-coherent states $M_Z=\pm S$ are semiclassical in 
nature, while the $M_Z=0$ state presents highly nonclassical features. 
This is somehow reflected in the linear (also referred to as 
{\it shot noise}) and in the quadratic (or {\it Heisenberg}) scaling with 
the system size $S$ of the measurement accuracy allowed by the two kinds 
of states \cite{Toth_2014a}. 

{\it Parameter estimation with a given observable. ---}
We have shown above that the spin projection along a suitable, temperature-dependent direction represents an optimal observable for precisely estimating the parameter $\lambda$. From a practical standpoint, it's also important to establish how robust such estimation protocol is, with respect to small deviations from optimality. More generally, one might ask what precision can be achieved by measuring a spin projection $ S_O = \hat{\bf n}_O \cdot {\bf S} $ along an arbitrary direction
$ \hat{\bf n}_O = [\sin(\theta+\phi) , 0 , \cos(\theta+\phi)] $. 
In order to answer these questions, we compute hereafter the classical Fisher information of $S_O$:
$
F(\rho_\lambda, S_O) 
= \sum_{M_O=-S}^S (\partial_\lambda p_{M_O})^2/p_{M_O} 
$ \cite{Helstrom_1976a},
where,
the probabilities corresponding to the possible measurement outcomes are given by
$ p_{M_O} = \sum_{M_Z} p_{M_Z} \langle M_O | M_Z \rangle^2 $, with $|M_O\rangle$ the eigenstates of $S_O$ \cite{SI}. 
\begin{figure}
\begin{center}
\includegraphics[width=0.45\textwidth]{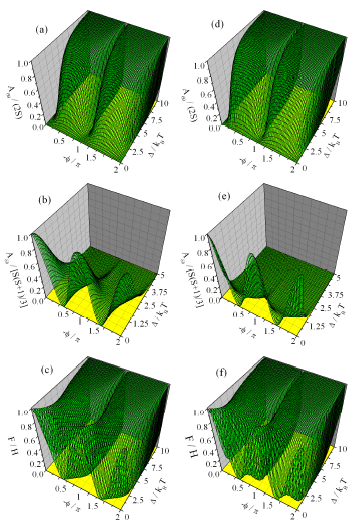}
\caption{\label{fig2}
Dependence of the Fisher information on the measured spin projection $S_O$ and on the ratio between Hamiltonian and thermal energy scales. The normalized terms $A_{\theta\theta}$ (a,d) and $A_{\delta\delta}$ (b,e) are the ones that enter the expression of $F$ given in Eq. (\ref{eqfia}). The overall value of the Fisher information is normalized to the quantum Fisher information (c,f), for $\dot\delta=\dot\theta$. The panels on the left-hand side refer to the case $S=1$, those on the right to $S=5$. 
}
\end{center}
\end{figure}
The classical Fisher information essentially quantifies the sensitivity of the probability distribution $\{p_{M_O}\}$ to the precise value of $\lambda$ and gives, through the Cramer-Rao bound, the highest precision that is achievable in a parameter estimation based on the measurement of the observable $S_O$.
\par
For the sake of the following discussion, we decompose $F$ into the sum of three independent contributions, according to the following expression:
\begin{equation}\label{eqfia}
F
= 
\sum_{\eta,\chi=\theta,\delta}
A_{\eta\chi} (\delta,\phi) \, \dot\eta \, \dot\chi .
\end{equation}
Here, $A_{\theta\theta}$ ($A_{\delta\delta}$) accounts for the dependence of $\rho_\lambda$ on the field orientation (intensity) alone, as reflected by the probabilities $p_{M_O}$. The term $A_{\delta\theta}=A_{\theta\delta}$, instead, results from the dependence of the equilibrium state on both $\Delta (\lambda)$ and $\theta (\lambda)$, and accounts for the interplay between the two. 
\par
As representative examples of the dependence of the classical Fisher information on $\delta=\Delta/k_BT$ and $\phi$, we consider the cases of two different spin lengths, namely $S=1$ and $S=5$ (Fig. \ref{fig2}). As can be seen from the plots (and from other cases \cite{SI}), $A_{\theta\theta}$ strongly increases with $\delta$, and achieves its maximum of $2S$ in the low-temperature limit (large $\delta$). As to the dependence on the observable $S_O$, this is very weak, apart from the dips corresponding to $\hat{\bf n}_O$ parallel ($\phi=0$) or antiparallel  ($\phi=\pi$) to the field orientation $\hat{\bf n}_Z$ [see panels (a,d)]. This contribution to the Fisher information, related to the dependence of the measurement statistics on the field orientation, is thus largely insensitive to deviations of the observable from optimality, especially for low temperatures and large values of $S$. The contribution $A_{\delta\delta}$, instead, increases for decreasing values of $\delta$, {\it i.e.} for high temperatures or small field intensities [panels (b,e)]. For $\delta \ll 1$ and $\phi=0,\pi$, such term achieves a maximum, whose value quite remarkably corresponds to $S(S+1)/3$, and thus scales quadratically with the spin length $S$. Besides, we note that the variation of $A_{\delta\delta}$ with $\delta$ becomes more rapid for larger values of $S$. The absolute value of the classical Fisher information contextually reflects the dependence of the equilibrium state on $\lambda$ and the capability of the observable $S_O$ to capture such dependence. In order to single out the latter effect and to assess the robustness of the estimation protocol with respect to small deviations of the observable $S_O$ from optimality, we finally consider the the ratio between the classical and the quantum Fisher information [panels (d,f)]. As clearly emerges from the plots, the value of $F/H$ remains close to the theoretical maximum of 1 in a wide region of the parameter space. This extends from the low- ($\delta \gg 1$) to the high-temperature regime ($\delta \ll 1$), where the classical Fisher information is dominated by the terms $A_{\theta\theta}$ and $A_{\delta\delta}$,respectively. 
\par
In the zero-temperature limit, where $\rho_\lambda$ coincides with the Hamiltonian ground state $|M_Z=-S\rangle$, the maximum precision in the parameter estimation can be obtained with arbitrary observables $S_O$. In fact, if the system is in an arbitrary eigenstate $|M_Z\rangle$ of $\mathcal{H}$, the expression of $F$ takes a simple analytical form:
\begin{equation}
F = \frac{(\hat{\bf n}_O \cdot \hat{\bf n}_X)^2}{(\hat{\bf n}_O \cdot \hat{\bf n}_X)^2+(\hat{\bf n}_O \cdot \hat{\bf n}_Y)^2}\, 2 [S(S+1)-M_Z^2] \dot\theta^2,
\end{equation}
where we have relaxed the assumption that $\hat{\bf n}_O$ lies in the $xz$ plane. From this, one can easily verify that $F=H$ whenever $\hat{\bf n}_O \perp \hat{\bf n}_Y$. Therefore, beyond what suggested by the zero-temperature limit of $L_\lambda$, any spin projection $S_O$ oriented along the $xz$ plane corresponds to an optimal observable is the system is prepared in an arbitrary state $|M_Z\rangle$, including the ground state $M_Z=-S$ of the Zeeman Hamiltonian. If instead $S_O$ has an out-of-plane component, the directions in the $xz$ plane are no longer equivalent, and only the component along $X$ contributes to the Fisher information. 

So far, we have considered the case where the parameter estimation is based on the outcome of a projective measurement, whose statistics is given by the probabilities $p_{M_O}$. In an ensemble measurement, the outcome is deterministically given by $\langle S_O \rangle = {\rm Tr} (\rho_\lambda S_O)$, from which one infers the value of $\lambda$, and is affected by an uncertainty Var$(S_O) = {\rm Tr} (\rho_\lambda S_O^2) - \langle S_O \rangle^2$. The precision in the parameter estimation can thus identified with the ratio 
$
P = |\partial_\lambda \langle S_O \rangle|^2 / {\rm Var} (S_O) 
$ \cite{Toth_2014a}. After expressing the above quantities in terms of the expectation values of $S_Z$ and $S_Z^2$, which are known functions of $\delta$, one obtains:
\begin{equation}
P 
\!=\! 
\frac{[\dot\theta\langle S_Z \rangle\sin \phi - \dot\delta \cos \phi \,
(\langle S_Z^2 \rangle \!-\! \langle S_Z \rangle^2)
]^2}{\langle S_X^2\rangle \sin^2 \phi \!+\! 
(\langle S_Z^2 \rangle \!-\! \langle S_Z \rangle^2)
\cos^2 \phi} \le F.
\end{equation}
\par
Interestingly, the numerical calculations show that the value of $P$ is either equal or slightly below that of $F$ throughout the parameter space \cite{SI}. Besides, one can show that $ P = F = H $ at any temperature for an optimal observable [corresponding to $\phi=\phi_{opt}$, see Eq. (\ref{eqOA})], and for arbitrary observables (i.e., for any value of $\phi$) in the low-temperature limit. In the qubit case ($S=1/2$), the precision $P$ can be expressed in a simple analytical form in terms of $\phi$ and $\theta$,
\begin{eqnarray}
P\!\!=\!F\!\!=\!\!\frac{\{\! 2\dot\theta\sin\phi\tanh(\delta\!/2)\!+\!\dot\delta\cos\phi[1\!-\!\!\tanh^2(\delta\!/2)]\}^2}{4\{\sin^2\phi\!+\!\cos^2\phi[1\!-\!\tanh^2(\delta/2)]\}} ,
\end{eqnarray}
and coincides with the classical Fisher information in all the parameter space (see Ref. \cite{Ghirardi_2017a} for further results on the qubit case). All this implies that an ensemble measurement allows parameter estimation with essentially the same precision as a (single) projective measurement. In the latter case, however, the precision can be increased, in principle indefinitely, by repeating the measurement many times. 

{\it Discussion and conclusions. ---}
The classical and the quantum Fisher information present contributions that scale either linearly or quadratically with the spin length. Relatively large values of $S$ can be achieved at the single-molecule level, in systems whose size is of the order of 1 nm \cite{Gatteschi_2007a,Troiani_2016a,Chen_2016a}. 
The total spin length can be increased (on average) by orders of magnitude, if one passes 
from single molecules to spin ensembles. The density operator of these systems can be written as a mixture of different contributions, each one corresponding to different a
value of $S$. In such a case, being the angle $\phi_{opt}$ independent on the spin length, the optimal observable $S_{O,opt}$ and the quantum Fisher information are equally well defined, and coincide with the expressions given above for the case of a specific spin length. Individual spins and spin ensembles thus offer two alternative routes to achieve precise parameter estimation: in the former case, where projective measurements can be implemented, one can exploit the repetition of the read out; in the latter case, where ensemble measurements are typically performed, one can benefit from the length of the collective spin.

In conclusion, we have considered the problem of sensing a magnetic field by measuring an arbitrary spin $S$ at equilibrium. The highest achievable precision, as quantified by the quantum Fisher information, can be expressed in terms of the spin magnetization and of the magnetic susceptibility, and, at finite-temperatures, results from a classical and a quantum contribution. These contributions can be related to the sensing of the field intensity and direction, respectively, they scale linearly and quadratically with the spin length, and are predominant in the high- and low-temperature limits. 
\par
The optimal observable is represented by the spin projection along a direction whose dependence on the temperature is universal (i.e. identical for all spin lengths), and is shown to allow a maximally precise parameter estimation also by means of ensemble measurements. This, along with the universal character of the optimality condition, has potential implications for the metrological use of many-spin systems, where $S$ is not a well defined quantum number but has a large average value. If the system is initialized in an arbitrary eigenstate of the Hamiltonian, the highest precision in the parameter estimation can be achieved by measuring a spin projection along an infinite set of optimal directions. Finally, the highest precision that is achievable in the field sensing by measuring a generic spin projection is close to that allowed by the optimal observable even for significant deviations from the optimal angle, thus proving the robustness of the scheme with respect to imperfections.
\par
This work has been supported by UniMI through the H2020 Transition 
Grant 15-6-3008000-625, by CINECA through the contract QCEXDCI and
by JSPS FY2017 program "Geometric foundation of quantum estimation". 
MGAP is member of GNFM-INdAM.

\end{document}